\documentclass[reprint]{JASA}

\usepackage{graphicx}% Include figure files
\graphicspath{{imgs/}}

\usepackage{array}

\usepackage{dcolumn}% Align table columns on decimal point
\usepackage{amsmath,amsfonts,amssymb}% popular packages from the American 
\usepackage{upgreek}
\usepackage[safe]{tipa}
\DeclareMathOperator{\var}{var}
\DeclareMathOperator*{\argmin}{argmin}
\DeclareMathOperator{\diag}{diag}

\allowdisplaybreaks

\usepackage{color} 
\usepackage[normalem]{ulem} % must have [normalem] so it won't overload \emph (to keep the emphasis in italics)

\newcommand{\R}[1]{{\mathbb R^{#1}}}

\newcommand{\thetaab}{{\boldsymbol{\uptheta}}}
\newcommand{\thetaabi}[1][\ell]{{\boldsymbol{\uptheta}_{#1}}}
\newcommand{\thetaphi}{{\boldsymbol{\upphi}}}

\newcommand{\Pc}{{\mathcal{P}_c}}
\newcommand{\thetaphihat}[1][*]{{\hat{\boldsymbol{\upphi}}^{#1}}}
\newcommand{\thetaabhat}[1][*]{{\hat{\boldsymbol{\uptheta}}^{#1}}}

\begin{document}

\preprint{AIP/123-QED}

\title[Time-varying harmonic models of voice signals]{Time-varying harmonic models for voice signal analysis}

\author{Takeshi Ikuma}
\email{tikuma@ieee.org}
\affiliation{Department of Otolaryngology--Head and Neck Surgery, Louisiana State University Health Sciences Center, New Orleans, LA 70112}
\author{Andrew J. McWhorter}
\affiliation{Department of Otolaryngology--Head and Neck Surgery, Louisiana State University Health Sciences Center, New Orleans, LA 70112}
\author{Lacey Adkins}
\affiliation{Department of Otolaryngology--Head and Neck Surgery, Louisiana State University Health Sciences Center, New Orleans, LA 70112}
\author{Melda Kunduk}
\affiliation{Department of Communication Disorders, Louisiana State University, Baton Rouge, LA 70803}

\date{\today}% It is always \today, today,

\editorinitials{Submitted}
\def\theDOI{/tba}

\begin{abstract}
	Assessment of voice signals has long been performed with the assumption of periodicity as this facilitates analysis. Near periodicity of normal voice signals makes short-time harmonic modeling an appealing choice to extract vocal feature parameters. For dysphonic voice, however, a fixed harmonic structure could be too constrained as it strictly enforces periodicity in the model. Slight variation in amplitude or frequency in the signal may cause the model to misrepresent the observed signal. To address these issues, this paper presents a time-varying harmonic model, which allows its fundamental frequency and harmonic amplitudes to be polynomial functions of time. The model decouples the slow deviations of frequency and amplitude from fast irregular vocal fold vibratory behaviors such as subharmonics and diplophonia. The time-varying model is shown to track the frequency and amplitude modulations present in voice with severe tremor. This reduces the sensitivity of the model-based harmonics-to-noise ratio measures to slow frequency and amplitude variations while maintaining its sensitivity to increase in turbulent noise or the presence of irregular vibration. Other uses of the model include the vocal tract filter estimation and the rates of frequency and intensity changes. These use cases are experimentally demonstrated along with the modeling accuracy.

	\vspace{3mm}
	This article has been submitted to The Journal of the Acoustical Society of America. After it is published, it will be found at \href{https://asa.scitation.org/journal/jas}{https://asa.scitation.org/journal/jas}.
\end{abstract}

% 43.70.Jt		Instrumentation and methodology for speech production research
% 43.70.Gr	Larynx anatomy and function; voice production characteristics 
% 43.70.Dn	Disordered speech 
%Use showkeys class option if keyword display desired
\maketitle

\section{Introduction}\label{sec:intro}

Sustained normal voice with a fixed pitch is generally considered to be a nearly periodic phenomenon \cite[type 1 signal;][]{titze1994}, and it naturally leads to the use of harmonic signal models to represent voice signals in, e.g., speech analysis and synthesis \cite{mcaulay1986,stylianou2001}, mucosal wave extraction \cite{jiang2008}, and highspeed videoendoscopy analysis \cite{ikuma2012}. Not limited to the model-based approaches, the periodicity assumption is the foundation of voice signal analyses, including perturbation analysis, spectral analysis, and glottal source analysis \cite{baken2000}. Under this nearly periodic (or quasi-stationarity) assumption, voice-related signals can be expressed as a Fourier series over a short analysis window:
\begin{equation} \label{eq:stationary}
	\frac{a_0}{2} + \sum_{p=1}^{\infty}{A_p \cos\left(2\pi p F_0 t + \Phi_p\right)} + v(t),
\end{equation}
where $F_0$ is the fundamental frequency in Hz, $a_0$ is the dc offset, $A_p$ is the magnitude coefficient of the $p$th harmonic, $\Phi_p$ is the phase coefficient for the $p$th harmonic in radians/second, and $v(t)$ represents the turbulent noise. The magnitude and phase parameters are slowly time-varying but the changes are assumed negligible during the window duration. 

For the stationary harmonic model in \eqref{eq:stationary} to describe dysphonic voices, aperiodic behaviors must be accounted by the term $v(t)$. It no longer represents only the turbulent noise but also the dysphonic effects like frequency modulation, amplitude modulation, subharmonics, and diplophonia \cite[type-2 and type-3 signals;][]{titze1994}. Many of the alterations are additive, i.e., the structure of the harmonics (if present) itself is unchanged but is contaminated by the added noise or nonharmonic tones. The additive effects do not necessarily hinder the analysis as the model could be processed further by decoupling the nonharmonic tones from $v(t)$ \cite{ikuma2013} or just treating $v(t)$ as overall noise. 

On the other hand, some dysphonic effect---such as frequency modulation and slow intensity fluctuation---violates the structural construct of the model because their modifications of the voice signals cannot be interpreted as additive. As such, the fitted harmonic terms poorly represent the actual vocal fold vibration. The frequency modulation especially causes the model to deviate from the signal as the model's cycle intervals do not match those of the signals. 

\begin{figure}
	\includegraphics{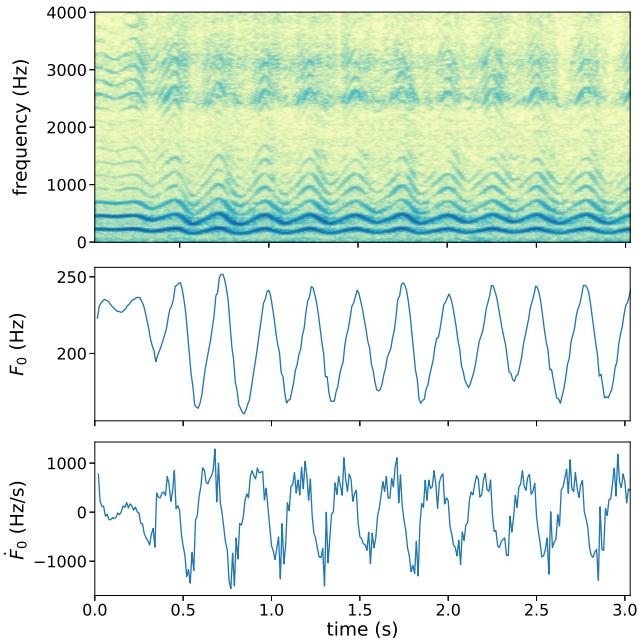}
	\caption{(color online) Spectrogram, fundamental frequency $F_0$, and the rate of $F_0$ change $F_{0,rate}$ of /i\textlengthmark/ phonation by a speaker with severe vocal fold tremor (sampled at $4000$ S/s, evaluated every $10$ ms). The rate of $F_0$ change is derived numerically from the $F_0$ measurements.}
	\label{fig_motiv}
\end{figure}

For example, vocal fold tremor is known to induce involuntary frequency and amplitude modulations in voice. Its severe form is shown in FIG. \ref{fig_motiv}.
% The observed maximum absolute rate of $F_0$ change is about $1200$ Hz/s, which translates to $60$-Hz (more than 25\% of the average $F_0$) shift over a $50$-ms window. 
Aside from the frequency modulation and, to a lesser degree, intensity fluctuation, the spectrogram appears normal: no apparent interharmonic activities suggesting irregular vocal fold vibration, no significant steepening of spectral tilt (i.e., no significant loss of intensities of the higher harmonics). There are no apparent sign of increased noise level. The latter two are evident by the smeared but strong presence of the harmonics in the second and third formants. The smearing of the higher harmonics is caused by the excessive frequency modulation. These observations lead to the expectation of high harmonics-to-noise ratio (HNR), which is the ratio of the periodic component to the aperiodic component of the voice signal. However, the stationary harmonic model yields low HNR measurements because its poor fit causes the time-varying aspects of the harmonics to be classified as noise. The HNR is generally perceived as a measure, relating to vocal roughness or breathiness \cite[][pp. 280-3]{baken2000} and thus the low HNR value in this tremor case can wrongly indicate "roughness" or "breathiness" in voice. This type of misidentification could also occur with natural fluctuations of frequency and intensity of normal voice (albeit to a lesser degree) and may lessen the effectiveness of the HNR as an acoustic correlate of breathiness or roughness.

To minimize the effect of the time-varying nature of harmonics to the voice signal analyses, one approach is to shorten the window duration, making a trade-off between the error due to time-variance and the error from reduced number of samples. Another is to select a robust nonparametric algorithm to compute the target parameters so that the measurements are insensitive to the variation. Yet another approach, which is the focus of this paper, is to use a parametric approach with a model which accounts for the variation. Specifically, the parametric model, referred to as the time-varying harmonic model, allows deterministic variation of the harmonic waveform both in the fundamental frequency $F_0$ and the Fourier coefficients, $a_0$, $A_p$, and $\Phi_p$ in Eq. \eqref{eq:stationary}. The proposed model is designed to decouple the slow (but noticeable) harmonic variations from the nonharmonic elements of voice signals (turbulent noise and irregular vibration).

The rest of the paper is organized as follows. The time-varying voice signal model is defined in Section \ref{sec:model_def} followed by the estimation of the model parameters in Section \ref{sec:model_est}. Section \ref{sec:uses} introduces several potential uses of the signal model: two types of HNRs, vocal tract filter estimation, and the rate-of-change parameters. In Section \ref{sec:examples}, the accuracy and consistency of the model are first presented followed by the application of the model to three distinctive voice signal cases.

\section{Time-Varying Harmonic Waveform Model}\label{sec:model_def}

A voice signal $x(t)$ over a short period can be expressed as the sum of nonstationary harmonic component $s(t)$ and aperiodic component $v(t)$:
\begin{equation} \label{eq:x_t}
	x(t) = s(t) + v(t).
\end{equation}
The aperiodic component $v(t)$ represents the turbulent noise for normal voice. It, however, could include additional tonal content which may be induced by irregular vibration of pathological vocal folds and not a part of the harmonic series. The nonstationary harmonic component $s(t)$ allows deterministic perturbations of its frequency and shape and takes the form:
\begin{equation} \label{eq:s_t_magphs}
	s(t) = \frac{a_0}{2} + \sum_{p=1}^{\infty}{A_p(t) \cos\left(p \Phi_0(t) + \Phi_p(t)\right)},
\end{equation}
where $\{a_0(t), A_p(t), \Phi_p(t): p=1,2,\ldots\}$ are the dc offset, magnitude coefficients, and common phase coefficients, respectively, as likewise defined for the stationary version of the model in Eq. \eqref{eq:stationary} but now dependent on time, and $\Phi_0(t)$ is the common phase term, also a function of time, which is originally $2 \pi F_0 t$ in Eq. \eqref{eq:stationary}. Accordingly, the instantaneous fundamental frequency is given by
\begin{equation} \label{eq:F0_t}
	F_0(t) = \frac{1}{2\pi} \frac{\partial}{\partial t}\Phi_0(t).
\end{equation}
The Fourier series expansion can also be expressed in the sine-cosine form:
\begin{equation} \label{eq:s_t}
	s(t) = \frac{a_0(t)}{2} + \sum_{p=1}^{\infty}{a_p(t) \cos\left(p \Phi_0(t) \right) + b_p(t) \sin\left(p \Phi_0(t) \right)},
\end{equation}
where $$a_p(t) = A_p(t) \cos \Phi_p(t) \text{ and } b_p(t)  = A_p(t) \sin \Phi_p(t).$$ This form is more algebraically tractable and used in the model estimation.

The time-dependent parameters $A_p(t)$, $\Phi_0(t)$, $a_p(t)$, and $b_p(t)$ in Eqs. \eqref{eq:s_t_magphs} and \eqref{eq:s_t} are ideally maintaining constant values but are likely to exhibit slow deviation over the analysis window. Although an increased level of deviation is expected for dysphonic voice, the mid-phonation deviations certainly occupy smaller bandwidths than the fundamental frequency so that they do not interfere with the general harmonic structure of $s(t)$. In the proposed model, the time-varying parameters are estimated by their Taylor polynomials (i.e., truncated Taylor series expansion) at the center time $t_0$ of an analysis window:
\begin{align}
	\Phi_0(t) & \approx \sum_{\ell=1}^{L_\Phi}\Phi_0^{(\ell)}(t_0)h_\ell(t-t_0),  \label{eq:phi_0_t}          \\
	a_p(t)    & \approx \sum_{\ell=0}^{L}a_p^{(\ell)}(t_0)h_\ell(t-t_0), \quad p=0,1,\ldots, \label{eq:a_p_t}
	\\
	\intertext{and}
	b_p(t)    & \approx \sum_{\ell=0}^{L}b_p^{(\ell)}(t_0)h_\ell(t-t_0), \quad p=1,2,\ldots, \label{eq:b_p_t}
\end{align}
where $L_\Phi, L$ are the polynomial degrees, $f^{(\ell)}(t_0)$ denotes the $\ell$th derivative of a function $f(t)$ evaluated at $t_0$, and
\begin{equation}\label{eq:h_i_t}
	h_\ell(t) \triangleq t^\ell/\ell!
\end{equation}
with $\ell!$ denoting the factorial of $\ell$. The common phase polynomial $\Phi_0(t)$ excludes the zeroth term as it is accounted by the Fourier coefficients $\Phi_p(t)$. The Taylor polynomial is an attractive option as its structure lends itself well to the problem. Assuming that the analysis window captures a several vocal cycles in $x(t)$, the crude overall shift of frequency or intensity from the beginning to the end of the window is modeled by the linear model, and each additional degree of freedom generally improves the fit without drastically changing the bandwidth of the model. 

\section{Least Squares Estimation of Model Parameters}\label{sec:model_est}

The goal is to estimate all the Taylor polynomial coefficients (the derivative terms) in Eqs. \eqref{eq:phi_0_t}-\eqref{eq:b_p_t} and the samples of $v(t)$ in Eq. \eqref{eq:x_t} given the samples of $x(t)$ and an initial $F_0$ estimate, which near the truth. First, assume $x(t)$ is the output of the ideal antialiasing filter and is sampled at the rate $F_s$ samples-per-second (S/s). Then, $v(t)$ is bandlimited to $F_s/2$ Hz, and $s(t)$ contains a finite number of harmonics.

Let $x_n \triangleq x(nT_s)$ be the sampled version of $x(t)$ where $T_s \triangleq 1/F_s$ is the sampling interval, and the model is estimated over a $N$-sample window (i.e., $T\triangleq N/F_s$ seconds long). Then, the voice signal model in discrete time is given by
\begin{equation}
	x_n = s_n + v_n, \quad n \in \{0, 1, \ldots, N-1\},
\end{equation}
where $v_n \triangleq v(nT_s)$, and 
\begin{equation} \label{eq:s_n}
	s_n = \frac{a_{0,n}}{2} + \sum_{p=1}^{P}{a_{p,n} \cos\left(p \phi_{0,n} \right) + b_{p,n} \sin\left(p \phi_{0,n} \right)}.
\end{equation}
Here, $P$ is the order of the highest non-zero harmonic present in $s_n$ and the time-normalized versions of the Taylor polynomials are defined as
\begin{align}
	\phi_{0,n} & = \sum_{\ell=1}^{L_\phi}\phi_{0,\ell}h_\ell(n-n_0), \label{eq:phi_0_n}           \\
	a_{p,n}    & = \sum_{\ell=0}^{L}a_{p,\ell}h_\ell(n-n_0), \quad p=0,1,\ldots, \label{eq:a_p_n}
	\\
	\intertext{and}
	b_{p,n}    & = \sum_{\ell=0}^{L}b_{p,\ell}h_\ell(n-n_0), \quad p=1,2,\ldots. \label{eq:b_p_n}
\end{align}
The time offset $n_0 \triangleq (N-1)/2$ sets the polynomial reference point to be at the middle of the window, and $h_\ell(t)$ is defined in Eq. \eqref{eq:h_i_t}.

The parameters to be estimated are the common phase parameters $\thetaphi \triangleq [\phi_{0,1} ~ \cdots, \phi_{0,L_\phi}] \in \R{L_\phi}$ and the amplitude parameters $\thetaab \in \R{(2P+1)(L+1)}$, which is a concatenation of $\{\thetaabi \triangleq [a_{0,\ell} ~ \cdots ~ a_{P,\ell} ~ b_{1,\ell} ~ \cdots ~ b_{P,\ell}]^T \in \R{2P+1}: \ell = 0, \ldots, L\}$. The operator $(\cdot)^T$ denotes the matrix transpose. These model parameters are estimated by minimizing the least squares criterion:
\begin{equation} \label{eq:cost}
	G(\thetaphi,\thetaab) = \frac{1}{2}\sum_{n=0}^{N-1} [\hat{s}_n(\thetaphi,\thetaab) - x_n]^2,
\end{equation}
where $\hat{s}_n(\thetaphi,\thetaab)$ is the estimated harmonic signal in the form of Eq. \eqref{eq:s_n} with the values of $\thetaphi$ and $\thetaab$.

The minimization of $G(\thetaphi,\thetaab)$ is subject to a bound constraint to limit the instantaneous harmonic frequencies $pF_0(t)$. Defining proper limits to the harmonic frequencies is critical. At minimum, the bound must be set to enforce all the harmonic frequencies of the solution to be positive and below the Nyquist frequency ($F_s/2$ Hz). A tighter $F_0(t)$ bound could also be used to keep the range of $F_0(t)$ close to the initial fixed estimate. Note that the bounds are defined for the normalized instantaneous fundamental frequency, $f_{0,n} \triangleq F_0(nT_s)/F_s$, which has the form:
\begin{equation}
	f_{0,n}(\thetaphi) = \frac{1}{2\pi} \sum_{\ell=1}^{L_\phi}\phi_{0,\ell}h_{\ell-1}(n-n_0), \label{eq:f0_n}
\end{equation}
following Eqs. \eqref{eq:F0_t}, \eqref{eq:phi_0_t}, and \eqref{eq:phi_0_n}.

Another constraint, limiting the rate of changes of parameters, could also be applied to the minimization. Limiting the rate is an effective way to limit the bandwidths of the perturbation models, thus preventing a model from overfitting the signal beyond the desired slow-changing trends. The maximum rate change could be imposed on the absolute rate of frequency change, $|\partial F_0 / \partial t|$, and the absolute rate of shape changes, $|\partial a_p(t)/\partial t|$ and $|\partial b_p(t)/\partial t|$. The latter may be applied to any of the harmonics ($p=0,1,\ldots,P$). Taking the derivative of Eqs. \eqref{eq:f0_n}, \eqref{eq:a_p_n}, and \eqref{eq:b_p_n} (assuming $n$ is continuous) yields
\begin{align}
	\dot f_{0,n}(\thetaphi) & \triangleq \frac{1}{2\pi} \sum_{\ell=2}^{L_\phi}\phi_{0,\ell}h_{\ell-2}(n-n_0), \label{eq:f0dot_n} \\
	\dot a_{p,n}(\thetaab)  & \triangleq \sum_{\ell=1}^{L}a_{p,\ell}h_{\ell-1}(n-n_0), \label{eq:adot_n}                         \\
	\intertext{and}
	\dot b_{p,n}(\thetaab)  & \triangleq \sum_{\ell=1}^{L}b_{p,\ell}h_{\ell-1}(n-n_0). \label{eq:bdot_n}
\end{align}

Putting everything together, the full constrained least squares problem in hand is
\begin{align}\label{eq:minprob}
	\underset{\thetaphi,\thetaab}{\text{minimize}} & \quad G(\thetaphi,\thetaab)                                                            \\
	\text{subject to}
	                                               & \quad \min_n f_{0,n}(\thetaphi) > f_{0,min},\notag                                     \\
	                                               & \quad \max_n f_{0,n}(\thetaphi) \le f_{0,max},\notag                                   \\
	(optional)                                     & \quad \max_n |\dot{f}_{0,n}(\thetaphi)| \le \dot{f}_{max},\notag                       \\
	(optional)                                     & \quad \max_n |\dot{a}_{p,n}(\thetaab)| \le r_{max}, p \in \Pc, \text{and} \notag    \\
	(optional)                                     & \quad \max_n |\dot{b}_{p,n}(\thetaab)| \le r_{max}, p \in \Pc \setminus \{0\}\notag
\end{align}
where $f_{0,min}$ is the non-negative minimum $f_{0,n}$ (default: $0$), $f_{0,max}$ is the maximum $f_{0,n}$ (default: $0.5/P$), $\dot{f}_{max}$ is the maximum allowable rate of $f_{0,n}$ change, $r_{max}$ is the maximum allowable rate of changes of $a_{p,n}$ and $b_{p,n}$, and $\Pc$ is the set of harmonic indices to apply the constraints to. These constraints simplify to one or two linear constraint for low-order parameter models: one if the constraining function is a constant, or two if the function is a linear function so that only the extreme $n$ values (i.e., $n=0$ and $N-1$) need to be checked.

This minimization problem does not have a closed-form solution; thus, it must be numerically solved with a non-linear optimization algorithm. To reduce the complexity of the problem and to keep the $F_0$ estimate near the initial estimate, an iterative alternating algorithm is employed to solve this problem. Two sets of parameters, $\thetaphi$ and $\thetaab$ are estimated alternately by freezing the other set. On the $k$th iteration, the amplitude parameter estimates $\thetaabhat[]$ are updated first based on the previous common phase parameter estimates:
\begin{align}\label{eq:minprob_ab}
	\thetaabhat[(k)] = & \argmin_\thetaab G\left(\thetaphihat[(k-1)],\thetaab\right)                             \\
	\text{subject to}  & \notag                                                                                  \\
	(optional)         & \quad \max_n |\dot{a}_{p,n}(\thetaab)| \le r_{max}, p \in \Pc, \text{and} \notag     \\
	(optional)         & \quad \max_n |\dot{b}_{p,n}(\thetaab)| \le r_{max}, p \in \Pc \setminus \{0\}.\notag
\end{align}
It is followed by update of the common phase parameters:
\begin{align}\label{eq:minprob_phi}
	\thetaphihat[(k)] = & \argmin_\thetaphi G\left(\thetaphi,\thetaabhat[(k)]\right)       \\
	\text{subject to}
	                    & \quad \min_n f_{0,n}(\thetaphi) > f_{0,min},\notag               \\
	                    & \quad \max_n f_{0,n}(\thetaphi) \le f_{0,max},\notag             \\
	(optional)          & \quad \max_n |\dot{f}_{0,n}(\thetaphi)| \le \dot{f}_{max}.\notag
\end{align}
The iterations to continue until the change in $\thetaphihat[ ]$ reaches the tolerance level $\rho$:
\begin{equation}
	\left\|\thetaphihat[(k)]-\thetaphihat[(k-1)]\right\|^2 < \rho.
\end{equation}
The subproblems, \eqref{eq:minprob_ab} and \eqref{eq:minprob_phi}, could be solved with any numerical optimization algorithm, which supports the objective and constraint functions. For the results in this paper, the rate constraints are not imposed, and the unconstrained amplitude subproblem \eqref{eq:minprob_ab} has the closed-form solution as described in Appendix \ref{apdx:ampest}. The other subproblem \eqref{eq:minprob_phi} is solved numerically as described in Appendix \ref{apdx:phsest}.

% It is critical to obtain a good initial $F_0$ estimate because the surface of $G(\thetaphi,\thetaab)$ is highly multimodal, specifically in $\thetaphi$. On the other hand, given $\thetaphi$, $\thetaab$ has a closed-form solution (if $\thetaab$ is unconstrained). 

To initialize the parameters, it is imperative to obtain a reasonable estimate of $F_0$ of $x_n$ to set $\phi_{0,1}^{(0)}$. Here, we are assuming that a dysphonic $x_n$ would have some periodic elements. In the case of a pure aperiodic voice signal \cite[type-3 voice signal; ][]{titze1994}, the analysis returns $s_n=0$. The higher order $\phi_{0,\ell}$ are initialized to zero. Moreover, the first $L_\phi$ iterations of the common phase minimization \eqref{eq:minprob_phi} are performed on one $\phi_{0,\ell}$ at a time to minimize the risk of finding incorrect a local minimum. The solution is expected to be in the original convex region of $\phi_{0,1}^{(0)}$ and $\phi_{0,\ell}=0$.

Once the final parameter estimates, denoted as $\thetaphihat$ and $\thetaabhat$, are obtained, the estimated harmonic signal is given by $\hat{s}_n^* \triangleq \hat{s}_n(\thetaphihat, \thetaabhat)$ and the estimated combined disturbance signal is given by $\hat{v}_n \triangleq x_n - \hat{s}_n^*$. Finally, the continuous-time parameters---$\Phi_0^{(\ell)}$, $a_p^{(\ell)}$, $b_p^{(\ell)}$ in Eqs. \eqref{eq:phi_0_t}, \eqref{eq:a_p_t}, and \eqref{eq:b_p_t}, respectively--- are obtained by multiplying their discrete-time versions in $\thetaphihat$ and $\thetaabhat$ by $F_s^\ell$.

\section{Uses of Harmonic Models for Short-Time Analysis}\label{sec:uses}

The model-based voice (either acoustic or glottal) analysis is primarily targeted to supplement spectral analysis \cite[][Ch.7]{baken2000}, especially for short time windows (containing less than 10 vocal cycles). As the analysis window shortens, nonparametric spectral estimation methods (e.g., periodogram and spectrogram) suffer from excessive spectral leakage and loss of spectral resolution \cite{kay1981}. The mainlobes of harmonics widen to hide the noise floor or weak interharmonic tones. Even if interharmonic tones are visible, estimates of their frequency and power could be subject to bias. This is especially true for a male voice because its harmonics are closer together than a female voice. The harmonic model provides the structural skeleton to mitigate these shortcomings and to improve the analysis outcomes with limited number of data samples. This section presents three key voice features that the model-based analysis can be used: HNR, vocal tract filter estimation, and the rates of changes of harmonic characteristics.

% Also, the decoupling of the harmonics and aperiodic (turbulent noise and interharmonic tonal) components of observed voice signal is the advantage of the model-based analysis and is otherwise not possible with other analysis frameworks.

\subsection{Harmonics-to-Noise Ratio (HNR)}

The harmonics-to-noise ratio \cite{yumoto1982} and its kin (e.g., signal-to-noise ratio, \citealp{kojima1980}; normalized noise energy, \citealp{Kasuya1986}; and noise-to-harmonics ratio, \citealp{deliyski1993,kreiman2014}) form an important category of objective measures to identify the increase in interharmonic noise level relative to the harmonics and found to correlate with hoarseness. The HNR measure and the harmonic model are closely related as the model intends to separate the harmonics and noise components, and the HNR is their power ratio.

The model yields two types of HNRs: overall \cite[e.g.,][]{yumoto1982,boersma1993,qi1997,Kasuya1986} and frequency-band-specific \cite{dekrom1995,lively1970,childers1991}. Given an estimated harmonic model, the overall HNR of $x_n$ is estimated by
\begin{equation}\label{eq:hnr}
	\text{HNR} \triangleq \frac{P_s}{\var \hat{v}_n},
\end{equation}
where $\var\hat{v}_n$ denotes the sample variance of the aperiodic component $\hat{v}_n$ and the harmonic signal power is given by
\begin{equation}
	P_s \triangleq \sum_{n=1}^{P} H_p
\end{equation}
with the power of the $p$th harmonic,
\begin{equation}\label{eq:Hp_avg}
	H_p \triangleq \frac{1}{2N}\sum_{n=0}^{N-1}a_{p,n}^2(\thetaabhat{}) + b_{p,n}^2(\thetaabhat{}),
\end{equation}
where $a_{p,n}(\thetaabhat{})$ and $b_{p,n}(\thetaabhat{})$ are computed according to Eqs. \eqref{eq:a_p_n} and \eqref{eq:b_p_n}, respectively, with the parameters $\thetaabhat{}$. Here, the dc component (the zeroth harmonic with the $a_{0,n}$ coefficient) is excluded as it does not contribute to the voice. The HNRs are typically reported in decibels (dB, $10 \log(\text{HNR})$).

The frequency-band-specific HNRs over the frequency range $\mathcal{F} \triangleq [F_\text{start},F_\text{end}]$ can be computed with the discrete Fourier transform (DFT) of $\hat{v}_n$:
\begin{equation}\label{eq:Vk}
	\hat{V}_k = \sum_{n=0}^{N-1} {\hat{v}_n \exp\left(-j \frac{2\pi k n}{K}\right)},
\end{equation}
where $j=\sqrt{-1}$ and $K$ is the number of DFT samples. Then,
\begin{equation}\label{eq:hnr_fb}
	\text{HNR}(\mathcal{F}) \triangleq NK \frac{\sum_{p\in\mathcal{P}_{\mathcal{F}}}H_p}{\sum_{k \in \mathcal{K}_{\mathcal{F}}} |\hat{V}_k|^2},
\end{equation}
where $\mathcal{P}_{\mathcal{F}} \triangleq \{p \in \mathbb{N}: p \bar{F}_0 \in \mathcal{F}\}$ and $\mathcal{K}_{\mathcal{F}} \triangleq \{k \in \mathbb{Z}: kF_s/K \in \mathcal{F}\}$. Here $\bar{F}_0$ is the average fundamental frequency, which is given by
\begin{equation}
	\bar{F}_0 = \frac{F_s}{N}\sum_{n=0}^{N-1} f_{0,n}(\thetaphihat{}).
\end{equation}
By the Parseval's theorem, the two types of HNRs are identical if $\mathcal{F}=[0,F_s/2]$.

The spectrum of $\hat{v}_n$ in Eq. \eqref{eq:Vk} is computed without any special windowing function. Windowing is commonly used in existing frequency-domain based HNR estimation algorithms \cite{Kasuya1986,dekrom1993,qi1997}, in which non-rectangular windowing functions are applied to control the spectral leakage of finite-length signal. It is not critical in the computation of $\text{HNR}(\mathcal{F})$ so long as random noise is the predominant component of $\hat{v}_n$. If substantial interharmonic tones are present or expected, then an appropriate window function shall be introduced to Eq. \eqref{eq:Vk} to reduce the spectral leakage across the $\mathcal{F}$ boundary.

\subsection{Estimation of vocal tract filter}

An ability to estimate glottal source signals or their characteristics from acoustic signals is important to isolate glottal production from vocal tract effect. Specifically, glottal spectral tilt have been suggested as an indicator of voice types, e.g., normal, vocal fry, breathy, and falsetto \cite{childers1991}. To obtain glottal spectral tilt from acoustic signals, the vocal tract effect must be removed with a model of vocal tract filter, which is configured with the estimates of the formant frequencies and bandwidths \cite{kreiman2012}.

As noted by Kreiman et al. (\citeyear{kreiman2012}), the formant frequency estimation of acoustic signals via linear predictive coding (LPC) analysis can be problematic for signals with high spectral tilt or with high $F_0$. Their solution to this issue was incorporating the source measurement (the open quotient from highspeed videoendoscopy data) to improve the formant estimation. The harmonic model is a potential alternate solution to this issue without the need of secondary data.

The primary LPC analysis issue stems from attempting to estimate the frequency response of the vocal-tract filter from the spectral peak samples of vocal harmonics. Having too few peaks or too wide of peak separation prevents a reliable estimate of the vocal-tract filter. With the assumption that the vocal tract filters both harmonic source signal and turbulent noise equally, LPC analysis of the aperiodic component of the model, $\hat{v}_n$, could improve the formant analysis because $\hat{v}_n$ provides continuous support in frequency unlike the harmonics. Also, the resulting formant frequency estimates are likely more accurate because the turbulent noise is spectrally flatter---e.g., as modeled in \cite{samlan2011}---than source harmonics with negative spectral tilt. A followup study is necessary to test these hypothesis.

\subsection{Rates of frequency and power changes}

While the model is not designed to evaluate pitch-synchronous perturbation measures such as jitters and shimmers \cite{baken2000}, it can produce measures describing the trends in frequency and power based on their instantaneous rates of changes. These rates signify the degrees of slow frequency and amplitude modulation present in the observed signal. The (relatively) faster amplitude changes such as subharmonics are accounted by $v(t)$ as the extra tonal content.

Based on the instantaneous rate expressions (\ref{eq:f0dot_n}), (\ref{eq:adot_n}), and (\ref{eq:bdot_n}), the instantaneous change of the fundamental frequency is given by
\begin{equation}
	\dot{F}_{0,n} = F_s^2 \dot{f}_{0,n}(\thetaphihat{})
\end{equation}
in Hz/s, and the instantaneous change of the $p$th harmonic power is
\begin{equation}
	\dot{H}_{p,n} = F_s \left[a_n(\thetaabhat{}) \dot{a}_n(\thetaabhat{}) + b_n(\thetaabhat{}) \dot{b}_n(\thetaabhat{})\right]
\end{equation}
in 1/s. The latter could be combined to the instantaneous change of the total harmonic power by
\begin{equation}
	\dot{P}_{s,n} = \sum_{p=1}^P \dot{H}_{s,n}.
\end{equation}
Within each analysis window, various statistics could be calculated from these instantaneous values: e.g., the root-mean-square (rms) and the maximum absolute rate.

\section{Numerical Examples}\label{sec:examples}

\subsection{Assessment of modeling accuracy}

To understand the behavior of the time-varying harmonic model and its estimation algorithm, synthetic signals that exactly match the assumed structure are used as the measure of model estimation errors. Because the estimation error is bundled in $\hat{v}_n$ along with the aperiodic noise, the accuracy of the estimated models is quantified effectively by the HNR measurements. Unless otherwise noted, the base signal models used in this section are generated with the harmonic magnitudes $A_p = 2^{1-p}/\alpha$ and randomly drawn harmonic phases $\Phi_p \in (0.0,2\pi)$. The scaling factor $\alpha$ is introduced to normalize the harmonic power to one. The normalized fundamental frequency is either stationary, $f_0 = 0.03$, or with linear frequency modulation (LFM), $f_{0,n} = 0.03+0.00002n$. This corresponds to $F_0=150$ Hz and $\dot{F}_0=500$ Hz/s at the sampling rate of $5000$ samples-per-second (S/s). The number of harmonics depends on the fundamental frequency and is set to fill the discrete-time spectrum ($P=16$). The additive noise $w(t)$ is drawn from zero-mean Normal distribution with variance $\sigma^2=0.01$. This configuration yields the actual HNR of $20$ dB. Each case presented below modifies one parameter at a time to observe its effect on the HNRs. All the results are shown at $F_s = 5000$ S/s for ease of relating the results to voice signals.

The proposed model-based HNR was computed with two different configurations: the exact harmonic model with linear frequency term and constant magnitudes (labeled HM\textsubscript{2,0} as $L_\phi=2$ and $L=0$) and the fixed-frequency model (HM\textsubscript{1,0}). The latter is identical to the model presented in \cite{ikuma2012}. In addition, three other HNR estimators are included to indicate the estimation errors relative to other methods: autocorrelation-based method implemented in Praat software \cite{boersma1993}, Qi's pitch-synchronous time-domain method (Qi-TD, \citealt{qi1997}), and Qi's cepstrum-aided frequency-domain method (Qi-FD, \citealt{qi1997}). The default window size is $250$ samples (or $50$ ms at $F_s=5000$ S/s). Each algorithm was tested in Monte Carlo simulation (10,000 realizations) and the mean HNR values are shown. The proposed and Qi-FD methods are initialized with the known $F_0$ while Praat estimates the HNR as a part of its pitch analysis procedure, and Qi-TD rely on Praat's pitch and pitch cycle analyses.

\begin{figure}
	\includegraphics{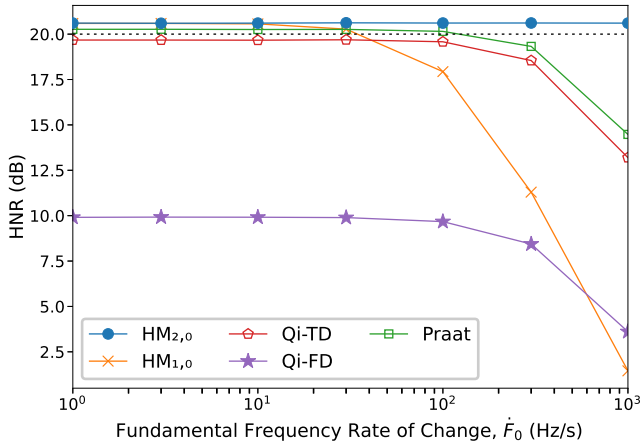}
	\caption{(color online) Rate of change of $F_0$, $\dot{F}_0$ vs. estimated HNR of synthetic signal (dotted line: actual HNR = 20 dB).}
	\label{fig02}
\end{figure}

FIG. \ref{fig02} shows the HNR estimates as $\dot{F}_0$ is varied. The proposed algorithm with correct model order, HM\textsubscript{2,0}, consistently reports 20.6 dB regardless of $\dot{F}_0$. It slightly overestimates the actual HNR due to overfitting to the noise component. The effect of fixing the $F_0$ in HM\textsubscript{1,0} is apparent as the HNR drops $\dot{F}_0>10$ Hz/s. All of the existing methods are more resilient to the $\dot{F}_0$ effect than HM\textsubscript{1,0} as they maintain their performance up to $\dot{F}_0=100$ Hz/s. This indicates the general lack of robustness of the model-based approach; deterministic nature of the harmonic model is more sensitive to model mismatch than nonparametric alternatives.

\begin{figure}
	\includegraphics{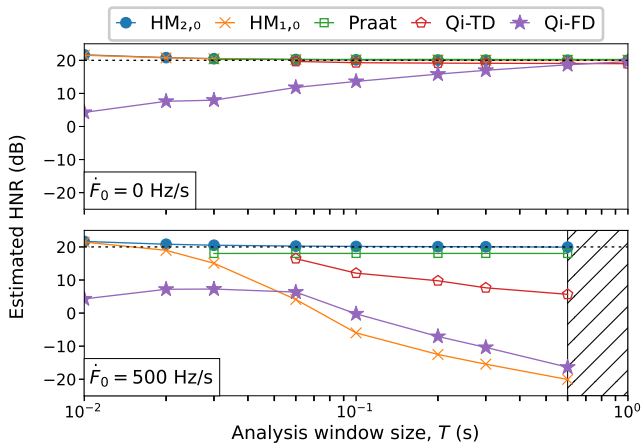}
	\caption{(color online) Analysis window size vs. estimated HNR of synthetic signals (dotted horizontal line: actual HNR = 20 dB). The ($\dot{F}_0=500$ Hz/s) cases are limited to $T<0.6$ s to maintain nonnegative instantaneous $F_0$ at all time.}
	\label{fig03}
\end{figure}

Qi-FD method severely underestimates the HNR with a large $-10$-dB bias at $\dot{F}_0=1$ Hz/s in FIG. \ref{fig02} while the others are within $1$ dB of the truth. This is caused by the spectral leakage effect \cite{qi1997} as it becomes apparent when the estimated HNR is shown as a function of the analysis window size $T$ in FIG. \ref{fig03}. Under $\dot{F}_0=0$, the Qi-FD estimates converge towards the actual HNR as $T$ increases although the trend reverses when the LFM is present, which also affects HM\textsubscript{1,0} and Qi-TD. The longer the window, the larger the $F_0$ difference at the beginning and end of the window due to the LFM, causing these algorithms to breakdown.
% This is due to the LFM continually increasing $F_0$ beyond realistic range for voice signals. 
On the other hand, HM\textsubscript{2,0} and Praat measurements are consistent under both $\dot{F}_0$ settings albeit Praat HNRs have slightly increased bias when the LFM is present. Note that Praat's inability to produce HNRs with window sizes $<30$ ms due to the pitch period is longer than the temporal support of the autocorrelation function at those window sizes.

\begin{figure}
	\includegraphics{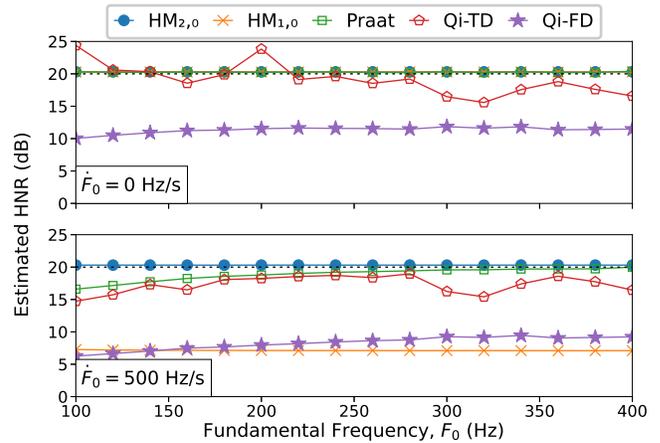}
	\caption{(color online) Average fundamental frequency vs. estimated HNR of synthetic signals (dotted horizontal line: actual HNR = 20 dB).}
	\label{fig04}
\end{figure}

\begin{figure}
	\includegraphics{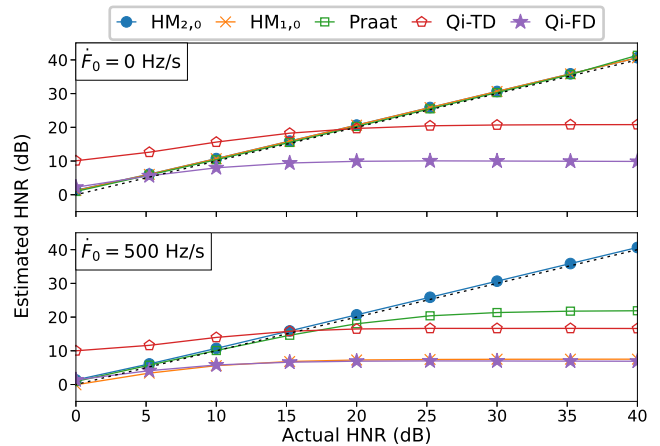}
	\caption{(color online) HNR vs. estimated HNR of synthetic signals (dotted diagonal line: ideal estimator).}
	\label{fig05}
\end{figure}

To complete the study of the HNR, FIG. \ref{fig04} illustrates the effect of the fundamental frequency, and FIG. \ref{fig05} shows the effect of the actual HNR (i.e., the noise power). The properly configured model (HM\textsubscript{2,0}) consistently demonstrates its ability to estimate the HNR regardless of the LFM presence. Praat's autocorrelation method also performs respectably, except that its robustness towards the LFM hinges on the $\dot{F}_0/F_0$ ratio.

Overall, the HM\textsubscript{2,0} model has consistently shown a computational error of less than 1 dB, which Praat with its autocorrelation-function-based algorithm matched (or slightly outperformed) for moderate LFM settings (up to 100 Hz/s). It shall be emphasized that the proposed method has significantly higher computational cost (Praat's estimation is near instantaneous while the Python implementation of the proposed algorithm takes 100 ms or longer, depending on the configuration). As an HNR estimator, HM\textsubscript{2,0} is useful for cases with the presence of strong LFM (above 100 Hz/s) where Praat loses its accuracy.

\subsection{Three case studies with voice signals}\label{sec:examples:cases}

While the results in the previous section have shown that the proposed least squares estimation yields well-fitted harmonic models with accurate HNR estimates when the input signal exactly follows the assumed structure, the actual voice signals do not completely adhere to the assumed structure. Both the amplitude and frequency variations of the harmonics are not polynomial, and the additive noise in acoustic signals is colored by the vocal tract effect. In this section, selected acoustic signals are modeled and analyzed using the model as suggested in Section \ref{sec:uses}. Three examples are one case each with vocal tremor and unilateral paralysis (collected by author MK) and one normal case in Saarbruecken Voice Database \cite{putzer2008}. The normal data include both acoustic and electroglottogram (EGG) waveforms.

\begin{figure}
	\includegraphics{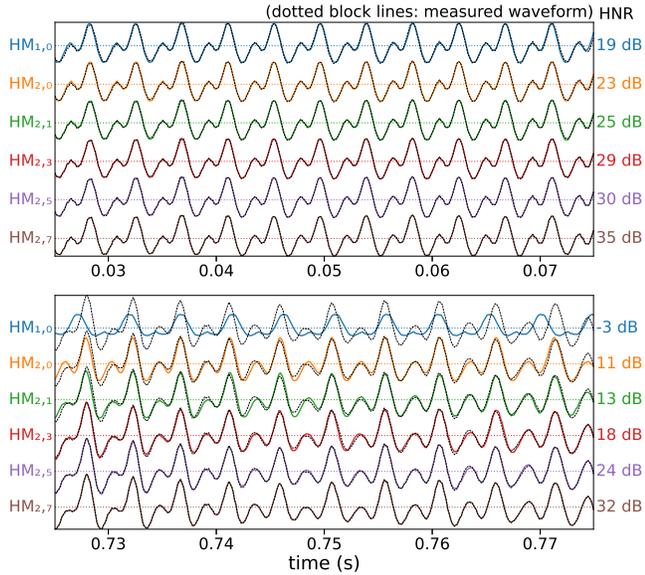}
	\caption{(color online) Fitted signals $\hat{s}(t)$ and their HNR estimates of two distinctive 50-ms windows of the tremulous signal in FIG. \ref{fig_motiv} with six model configurations (HM$_{L_\phi,L}$): (top) at $t=0.05$ when the tremor is minimally present, and (bottom) at $t=0.75$ when the intra-frame $F_0$ change is the largest. ($F_s=4000$ S/s, $T=0.05$ s, $P=8$, initial $F_0$ provided by Praat)}
	\label{fig06}
\end{figure}

First, the vocal fold tremor case in FIG. \ref{fig_motiv} is revisited. FIG. \ref{fig06} illustrates the signal modeling outcomes at two distinctive times.
The most basic fixed model, HM\textsubscript{1,0}, already fits well to the signal at $t=0.05$ with the estimated HNR of 19 dB. Inclusion of linear frequency modulation to the model (i.e., HM\textsubscript{2,0}) minimally increased the estimated HNR by 4 dB, mainly by correcting the misfits near the edges of the analysis window. The HNR growth with the magnitude polynomial order is steady but small. In contrast, HM\textsubscript{1,0} severely misfits the signal at $t=0.75$ due to the presence of substantial frequency modulation, and the misalignment of the pitch cycles in the measured and modeled signals is visibly recognizable. Using the LFM-enabled model, HM\textsubscript{2,0}, largely fixes the cycle alignment (thereby increasing the HNR by 14 dB), but it still does not account for the slow amplitude modulation present in the signal. Increasing $L$ gradually improves the model to match the amplitude behavior.

\begin{figure}
	\includegraphics{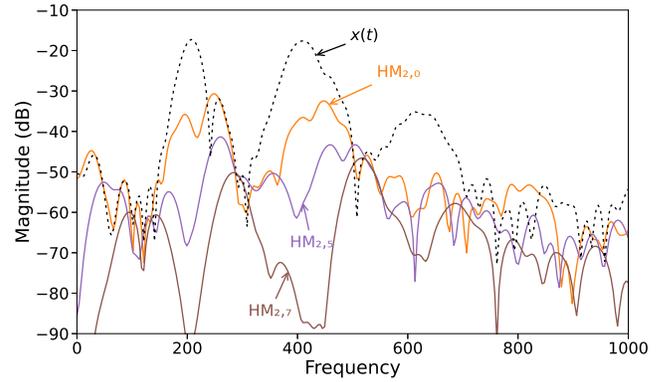}
	\caption{(color online) Spectra of estimated aperiodic noise $\hat{v}(t)$ of the selected fitted signal $x(t)$ from $t=0.75$ case in FIG. \ref{fig06}. The modeled spectra are computed from three different models, annotated as HM$_{L_\phi,L}$}.
	\label{fig07}
\end{figure}

The fit of the model is better visualized in the frequency domain by inspecting the spectrum of the residual $\hat{v}(t)$ over the first 3 harmonics as shown in FIG. \ref{fig07}. 
Ideally, $\hat{v}(t)$ contains only the turbulent noise and interharmonic tones and no harmonics. The HM\textsubscript{2,0} model substantially reduces the harmonic peaks of $x(t)$ but not entirely. HM\textsubscript{2,7}, on the other hand, overfits the signal as indicated by the formation of valleys at the harmonic frequencies. The HM\textsubscript{2,5} model best separates the harmonics from the noise. It also reveals the presence of weak amplitude modulation. The extraneous tone is apparent by the first harmonic in $x(t)$ but another one is buried in the mainlobes of the second harmonics. Arbitrarily increasing the polynomial order causes the model to overfit the signal and becomes a detriment to the analysis. There are two approaches to the model order selection. First is to use the lowest order, which is unlikely to cause an overfit but may underfit extreme cases like the tremor case under study. Another approach is to employ the information theoretic criterions such as Akaike information criterion or minimum description length \cite{wax1985} to pick the optimal orders automatically. Further investigation is necessary to establish the model order selection strategy.

\begin{figure}
	\includegraphics{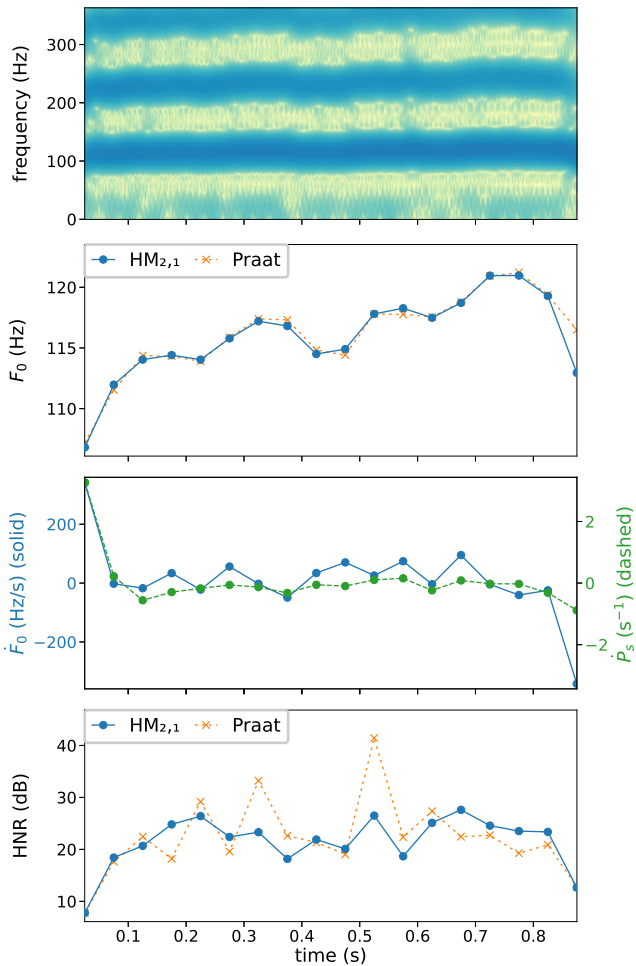}
	\caption{(color online) HM\textsubscript{2,1}-based analysis outcomes of /a\textlengthmark/ vowel of normophonic speaker: narrowband spectrogram, $F_0$ estimates, most extreme $\dot{F}_0$ (change of instantaneous $F_0$) and $\dot{P}_s$ (rate of instantaneous harmonic power), and HNR. Praat's $F_0$ and HNR estimates are also shown. ($F_s=8000$ S/s, $T=50$ ms, $L_\phi=2$, $L=1$)}
	\label{fig08}
\end{figure}

\begin{figure}
	\includegraphics{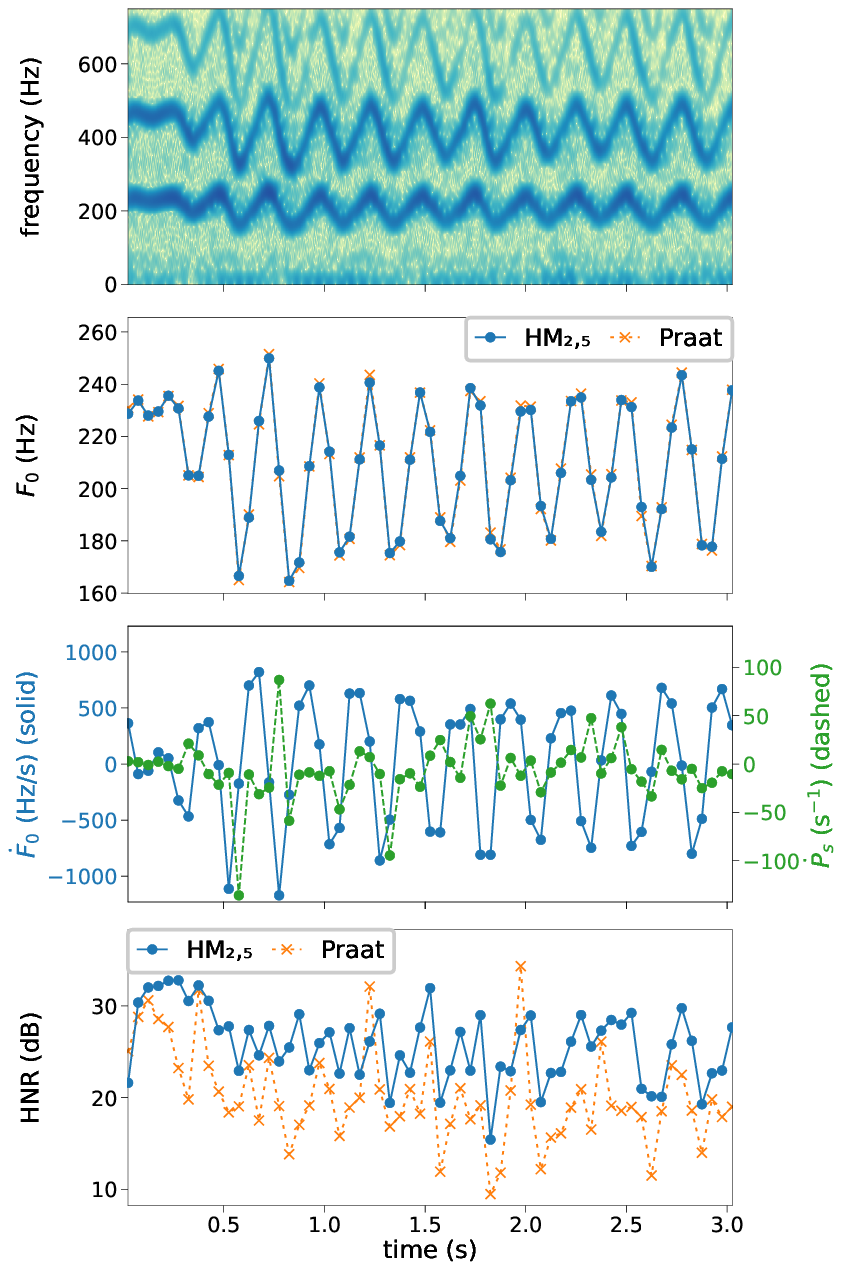}
	\caption{(color online) HM\textsubscript{2,5}-based analysis outcomes of /i\textlengthmark/ vowel of speaker with tremor in FIG. \ref{fig_motiv}: narrowband spectrogram, $F_0$ estimates, most extreme $\dot{F}_0$ (change of instantaneous $F_0$) and $\dot{P}_s$ (rate of instantaneous harmonic power), and HNR. Praat's $F_0$ and HNR estimates are also shown. ($F_s=8000$ S/s, $T=50$ ms, $L_\phi=2$, $L=5$)}
	\label{fig09}
\end{figure}

\begin{figure}
	\includegraphics{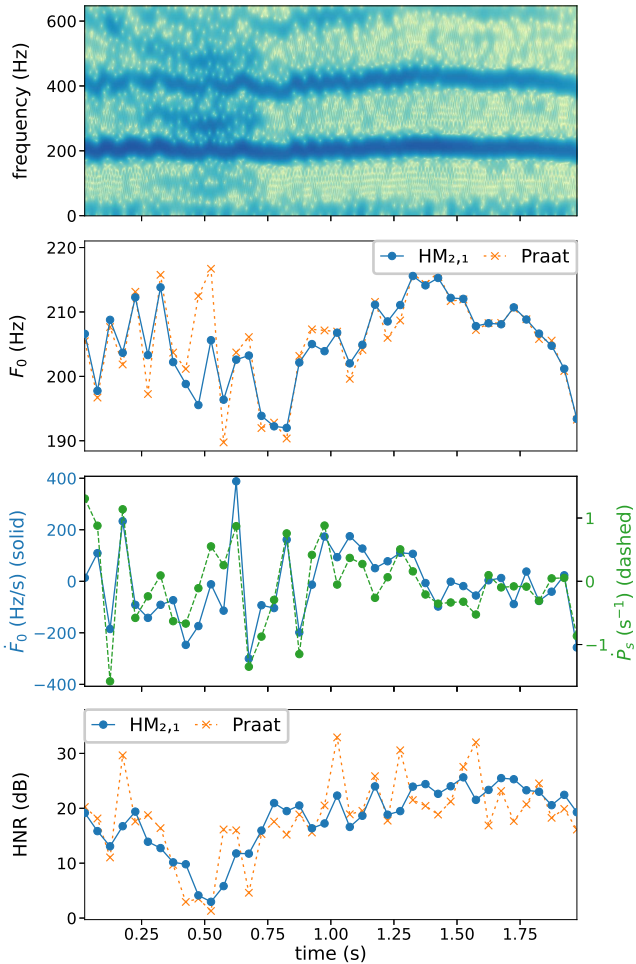}
	\caption{(color online) HM\textsubscript{2,3}-based analysis outcomes of /i\textlengthmark/ vowel of speaker with unilateral vocal fold paralysis: narrowband spectrogram, $F_0$ estimates, the most extreme $\dot{F}_0$ (change of instantaneous $F_0$) and $\dot{P}_s$ (rate of instantaneous harmonic power), and HNR. Praat's $F_0$ and HNR estimates are also shown. ($F_s=8000$ S/s, $T=50$ ms, $L_\phi=2$, $L=1$)}
	\label{fig10}
\end{figure}

Based on the manual model-order selection, the analysis outcomes of three distinctive cases are shown as follows: a normophonic male voice in FIG. \ref{fig08}, a female voice with vocal fold tremor in FIG. \ref{fig09}, and a female voice with unilateral vocal fold paralysis in FIG. \ref{fig10}. The model fitting was initialized with the Praat's $F_0$ estimate and $\dot{F}_0=0$, and the harmonic order was set for each case to fill the discrete-time frequency spectrum.

The HNR and $F_0$ of the normophonic case in FIG. \ref{fig08} confirms that model-based estimates closely follow those of Praat when the voice signal adheres to the harmonic signal model. The model-based HNR estimates across windows appear to have less fluctuation than the Praat's HNRs. This case also provides references for the rates of changes of instantaneous $F_0$ and $P_s$. In the middle of the phonation, $|\dot{F}_0|<100$ Hz/s and $|\dot{P}_s|<1.0$ s\textsuperscript{-1}. 

Praat and the harmonic model are also matched well on the overall variation of the instantaneous $F_0$ of the tremor case in FIG. \ref{fig09}, and the range of the estimated  $\dot{F}_0$ agrees with the inter-window $\dot{F}_0$ estimates in FIG. \ref{fig_motiv}. The range of $\dot{F}_0$ is an order of magnitude more than that of the normal case. The tremor is also prevalent in the intensity as $\dot{P}_s$ shows frequent spikes although its periodicity is not easily perceivable as the frequency. The model-based HNR measures are consistently higher than those measured by Praat. The range of the model-based HNRs is similar to the normal case while Praat HNRs drops below 10 dB when severe tremor is present. These observations are inline with the synthetic results in FIG. \ref{fig02}.

\begin{figure}
	\includegraphics{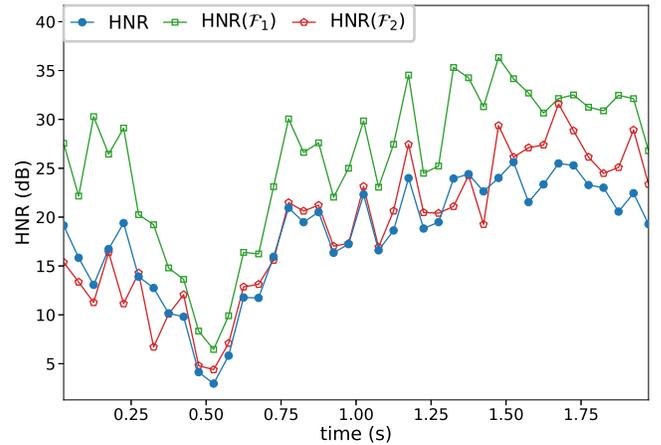}
	\caption{(color online) Overall and per-harmonic HNRs of the unilateral vocal fold paralysis case in FIG. \ref{fig10}: HNR=overall, HNR($\mathcal{F}_1$)=first harmonic HNR, HNR($\mathcal{F}_2$)=second harmonic HNR.}
	\label{fig11}
\end{figure}

The HNR is also known to be sensitive to turbulent noise and relates to a breathy quality of voice, specifically attributed to high frequency noise \cite{dekrom1995}. On the other hand, the irregular vocal fold vibrations are most prevalent near the most dominant harmonics often near the lower formant frequencies. As such, the per-harmonic HNR per \eqref{eq:hnr_fb}, inspired by Childers and Lee's NHR \citeyear{childers1991}, is more effective than the overall HNR to identify the analysis windows with irregular vibration as shown in FIG. \ref{fig11}. Here, $p$th-harmonic HNR is a frequency-specific HNR measure with the frequency range defined by $ \mathcal{F}_p = ((p-0.5)F_0, (p+0.5)F_0)$. The first harmonic HNR is about 10 dB higher during segments with normal vibration than the overall HNR (which represents 0-4 kHz) while the difference narrows to 5 dB on the most disturbed window.

\begin{figure}
	\includegraphics{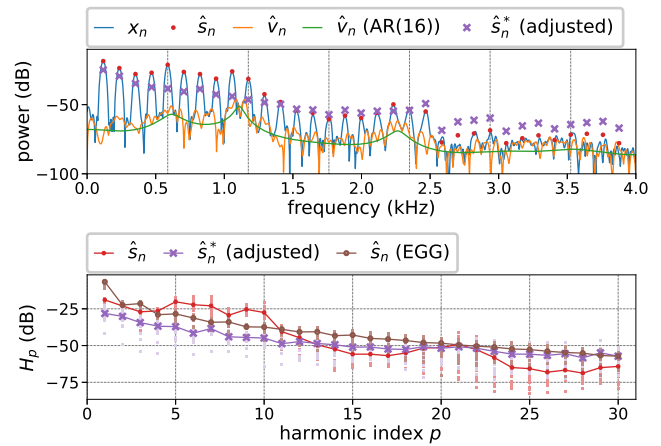}
	\caption{(color online) Vocal tract filter estimation of the normophonic case in FIG. \ref{fig08}: (top) estimated power spectra of signal $x_n$ and estimated disturbance $\hat{v}_n$, estimated harmonic peaks $\hat{s}_n$, AR(16) model of $\hat{v}_n$, and harmonic peaks after the filter effect is removed $\hat{s}_n^*$ in one analysis window. (bottom) series of harmonic power before and after the adjustment as well as those of simultaneously recorded EGG signal. The results are the averages over 18 windows, and small dots indicate the measurements from individual windows.}
	\label{fig12}
\end{figure}

Finally, the vocal tract filter estimation of the normophonic case is shown in FIG. \ref{fig12}. Here, the estimated noise $\hat{v}_n$ (which comprises mostly turbulent noise) underwent LPC analysis (Burg's method). The noise is modeled with an autoregressive (AR) model of order 16. The resulting AR(16) spectral peaks agrees with the formant peaks apparent with $x(t)$. The powers of the harmonic are then adjusted according to the associated AR filter. The resulting estimates of the source harmonics exhibit the expected monotonically descending trend of the glottal source signals, and its spectral tilt angle closely matches that of the simultaneously recorded EGG signal. While the efficacy of this approach has not been proven, the results in FIG. \ref{fig12} appear promising and encourage its use in evaluating formant frequencies and spectral tilt parameters.

\section{Conclusion}\label{sec:conc}

This paper presents the use of a time-varying harmonic model with deterministically time-varying fundamental frequency and amplitude parameters. These parameters are aimed to capture the slowly varying nature of voice as a part of the harmonic signal, instead of treating it as undesirable deviation. For the disordered voice, the model has shown its ability to separate disordered slow changes (e.g., amplitude and frequency modulations in tremor) from faster interharmonic irregular vocal fold vibration. The proposed model is useful in computing the frequency-specific HNRs with short analysis windows and enables measuring within-window rate of changes in instantaneous frequency and power. Finally, the model enables the direct estimation of vocal tract filter from the extracted turbulent noise.

%\begin{acknowledgments}
%We wish to acknowledge the support of the author community in using
%\JasaTeX{}, offering suggestions and encouragement, and testing new versions.
%\end{acknowledgments}

% ---------------------------------------------------------------------
% Appendix  (optional)

\appendix
\section{Amplitude parameter estimation}\label{apdx:ampest}

The minimization problem \eqref{eq:minprob_ab} is an (optionally constrained) linear least squares problem. If the constraints are omitted, this problem has the close-form solution as follows.

Substituing Eqs. \eqref{eq:a_p_n} and \eqref{eq:b_p_n} into Eq \eqref{eq:s_n} and subsequent algebraic manipulation yields
\begin{multline} \label{eq:s_n_alt}
	s_n(\thetaphi,\thetaab) = \sum_{\ell=0}^{L} h_\ell(n-n_0) \bigg[ \frac{1}{2}a_{0,\ell}
	+ \sum_{p=1}^{P} a_{p,\ell} \cos\left(p \phi_{0,n} \right) \\
	+ \sum_{p=1}^{P} b_{p,\ell} \sin\left(p \phi_{0,n} \right) \bigg]
\end{multline}
This can be rewritten in a vector-matrix format:
\begin{equation} \label{eq:s_n_alt2}
	\begin{split}
		\mathbf{s}(\thetaphi,\thetaab) &= \sum_{\ell=0}^{L} \mathbf{H}_\ell \mathbf{M}_\thetaphi \thetaabi[\ell] \\
		&= \mathbf{A}\left(\thetaphi\right) \thetaab,
	\end{split}
\end{equation}
where
\begin{align}
	\mathbf{s}            & \triangleq \left[\begin{matrix} s_0 & s_1 & \cdots & s_{N-1} \end{matrix}\right]^T                                                                                             \\
	\mathbf{H}_\ell       & \triangleq \diag ( h_\ell(-n_0), h_\ell(1-n_0), \ldots, \nonumber                                                                                                   \\
	                      & \omit \hfill $h_\ell(N-1-n_0))$,                                                                                                                                    \\
	\mathbf{M}_\thetaphi  & \triangleq
	\left[
		\begin{array} {c|c|c}
			\frac{1}{2}\mathbf{1} & \mathbf{C}\left(\thetaphi\right) & \mathbf{S}\left(\thetaphi\right)
		\end{array}
	\right].\label{eq:M}                                                                                                                                                                        \\
	\intertext{and}
	\mathbf{A}(\thetaphi) & = \left[\begin{array}{c|c|c|c|c}
			                                \mathbf{M}_\thetaphi & \mathbf{H}_1 \mathbf{M}_\thetaphi & \mathbf{H}_2 \mathbf{M}_\thetaphi & \cdots & \mathbf{H}_L \mathbf{M}_\thetaphi\end{array}\right]
\end{align}
Here, $\diag()$ defines a diagonal matrix with the specified diagonal elements, $\mathbf{1}$ is a vector of $N$ ones, and $\mathbf{C}(\thetaphi)$ and $\mathbf{S}(\thetaphi)$ are $(N \times P)$ matrices with the $np$-th element given by $\cos(p \phi_{0,n})$ and $\sin(p \phi_{0,n})$, respectively. The common phase $\phi_{0,n}$ is evaluated by Eq. \eqref{eq:phi_0_n} with the current values in $\thetaphi$. The optimal solution is then obtained by
\begin{equation}\label{eq:p_est}
	\thetaabhat[(k)] =
	\mathbf{A}^+\left(\thetaphihat[(k-1)]\right)
	\mathbf{s},
\end{equation}
where $^+$ denotes the matrix pseudo-inverse.

\section{Common phase parameter estimation}\label{apdx:phsest}

The minimization problem \eqref{eq:minprob_phi} is solved with the trust-region constrained algorithm \cite{byrd1999} implemented by Python SciPy package (v1.7.3, https://scipy.org). The Jacobian and Hessian matrix of the objective function are given as follows.

\begin{equation}
	\mathbf{J}  =
	\begin{bmatrix}
		\frac{\partial G}{\partial \phi_{0,1}} & \frac{\partial G}{\partial \phi_{0,2}} & \cdots & \frac{\partial G}{ \partial \phi_{0,L_\phi}}
	\end{bmatrix},
\end{equation}
where
\begin{equation}
	\frac{\partial G}{\partial \phi_{0,\ell}} = \sum_{n=0}^{N-1} e_n c_{1,n} h_\ell(n-n_0)
\end{equation}
with
\begin{align}
	e_n     & \triangleq \hat{s}_n - x_n                                                              \\
	\intertext{and}
	c_{1,n} & \triangleq \sum_{p=1}^{P} p \left[-a_{p,n}\sin (p\phi_n) + b_{p,n}\cos(p\phi_n)\right].
\end{align}
Samples of $\hat{s}_n$, $a_{p,n}$, and $b_{p,n}$ are evaluated with the current $\thetaphihat[ ]$ and $\thetaabhat[ ]$. The $ij$th element of the Hessian matrix $\mathbf{H} \in \R{L_\phi \times L_\phi}$ takes the form
\begin{align}
	\mathbf{H}|_{ij} & = \frac{\partial^2 G}{\partial \phi_{0,\ell} \partial \phi_{0,j}}                       \\
	                 & = \sum_{n=0}^{N-1} \left(c_{1,n}^2 + e_n c_{2,n}\right) h_\ell(n-n_0) h_j(n-n_0),\notag
\end{align}
where
\begin{equation}
	c_{2,n} \triangleq \sum_{p=1}^P p^2 \left[ a_{p,n} \cos (p\phi_n) + b_{p,n} \sin(p\phi_n) \right].
\end{equation}

%The following command formats the BiBTeX-generated bibliography by reading in the .bbl file.
%When preparing a TeX document for submission to the JASA, you must paste in the content
%of that file in place of this command: the Journal requires submission of a single .tex file.
%\bibliographystyle{plainnat}

\bibliography{Manuscript}%

\end{document}